\documentclass{aa}  
\usepackage{natbib}
\bibpunct{(}{)}{;}{a}{}{,} 
\usepackage{graphicx}
\usepackage{float}
\usepackage{xcolor}
\usepackage{graphicx}
\usepackage{dcolumn}
\usepackage{bm}
\usepackage{romannum}
\usepackage{amsmath}
\usepackage{graphicx}
\usepackage{xcolor}
\usepackage[caption=false]{subfig}
\usepackage{booktabs}   
\usepackage{relsize}
\usepackage{sidecap}

\usepackage{txfonts}

\begin{document}

\title{Infalling ultra-faint dwarfs as emissaries of the Axiverse}

   \author{A. Pozo
          \inst{1}
          \and
          T. Broadhurst\inst{1,2,7}
          \and
          H. N. Luu\inst{1}
          \and
          G. Smoot\inst{1,8,9,10,11}
          \and
          K. Umetsu\inst{3}
          \and
          T.  Chiueh\inst{4}
          \and
          H.-Y. Schive\inst{4}
          \and
          R. Emami\inst{5}
          \and
          L. Hernquist\inst{5}
          \and
          P. Mocz\inst{6}
          \and
          M. Vogelsberger\inst{12}     
          }

  \institute{DIPC, Basque Country UPV/EHU, E-48080 San Sebastian, Spain\\
              \email{alvaro.pozolarrocha@bizkaia.eu; tom.j.broadhurst@gmail.com;}
         \and
             University of the Basque Country UPV/EHU,Department of Theoretical Physics, Bilbao, E-48080, Spain
         \and
             ASIAA, Taipei 10617, Taiwan
         \and
             Department of Physics, National Taiwan University, Taipei 10617, Taiwan
         \and
             Center for Astrophysics $\vert$ Harvard $\&$ Smithsonian, 60 Garden Street, Cambridge, MA 02138, USA
          \and
             Center for Computational Astrophysics, Flatiron Institute, 162 5th Ave, New York, NY 10010, USA
         \and
             Ikerbasque, Basque Foundation for Science, Bilbao, E-48011, Spain
         \and
             Hong Kong University of Science and Technology, Institute for Advanced Study and Department of Physics, IAS TT $\&$ WF Chao Foundation Professor,  Hong Kong
         \and
             Energetic Cosmos Laboratory, Nazarbayev University, Nursultan, Kazakhstan
         \and
             Paris Centre for Cosmological Physics, APC, AstroParticule et Cosmologie, Universit\'{e} de Paris, CNRS/IN2P3, CEA/lrfu,Universit\'{e} Sorbonne Paris Cit\'{e}, 10, rue Alice Domon et Leonie Duquet,  75205 Paris CEDEX 13, France  Emeritus
        \and
             Physics Department $\&$ LBNL, University of California at  Berkeley CA 94720 {\it Emeritus}
        \and
             Department of Physics, Kavli Institute for Astrophysics and Space Research, Massachusetts Institute of Technology, Cambridge, MA 02139, USA
             }

  \abstract{
 Recent discoveries of ultra-faint dwarf galaxies (UFDs) infalling onto the Milky Way, namely Leo K \& M at $r \simeq 450$kpc, considerably strengthens the case that UFDs constitute a distinct galaxy class that is inherently smaller and fainter, and metal-poorer than the classical dwarf spheroidals (dSph). This distinction is at odds with the inherent continuity of galaxy halo masses formed under scale-free gravity for any standard dark-matter (DM) model. Here, we show that distinct galaxy classes do evolve in cosmological simulations of multiple light bosons representing the ¨Axiverse¨ proposal of string theory, where a discrete mass spectrum of axions is generically predicted to span many decades in mass. In this context, the
 observed UFD class we show corresponds to a relatively heavy boson of $3\times 10^{-21}$eV, including Leo K \& M, whereas a lighter axion of $10^{-22}$eV comprises the bulk of DM in all larger galaxies including the dSphs. Although Leo M is larger in size than Leo K, we predict its velocity dispersion to be smaller $(\simeq 1.7$km/s) than that of Leo K $(\simeq 4.5$km/s) because of the inverse de Broglie scale dependence on momentum. This scenario can be definitively tested using millisecond pulsars close to the Galactic center, where the Compton frequencies of the heavy and light bosons imprint monotone timing residuals that may be detected by the Square Kilometre Array (SKA) on timescales of approximately one week and four months, respectively.}


   \maketitle

\section{Introduction}

The increasingly well established class of ultra-faint dwarf galaxies (UFDs) \citep{Simon:2015,Simon:2019,Fritz:2018,McQuinn:2024} challenges all standardly explored dark-matter (DM) models, as the extreme properties of UFDs are not continuous with the general galaxy population, including the classical dwarf spheroidal (dSph) from which UFDs are readily distinguished with little overlap in terms of size, mass, luminosity, and metallicity. To date, over 40 galaxies have been classified as UFDs and were uncovered systematically in panoramic sky surveys, including the initial DES-related discoveries \citep{Simon:2015} that have been aided by Gaia astrometry and deep follow-up spectroscopy \citep{Fritz:2018}. These UFDs are found orbiting the Milky Way, typically well within the virial radius, and nearly all show no obvious evidence of tidal effects. In general, for those UFDs with sufficient stellar-velocity measurements, it has become clear that the mass distribution is dominated by DM, with notably higher mass-to-light ratios ($M/L$) than dSph galaxies by over an order of magnitude. UFD sizes are also smaller than those of dSphs by a factor of 5–10 and of much lower luminosity, typically $3 \times 10^4 L_{\odot}$. Furthermore, UFDs are also distinguishable in terms of metallicity \citep{Fu:2023}, with typical [Fe/H]s $\simeq -2.5$ compared to [Fe/H] $\simeq -1.7$ for dSph galaxies.

The relatively high DM density of UFDs can account for the lack of tidal stripping, with stellar orbits sufficiently deep in the potential. Nevertheless, the possibility of stripped stars has been debated, with upper limits on the density of such stellar halos estimated. Indeed, some UFDs are claimed to possess extended low-density outer halos (e.g., Tucana II \citep{Chiti:2021,Chiti:2022}) established with spectroscopy; for this UFD, tidal stripping is not claimed to be significant. Other such halos are tentatively claimed and may plausibly emerge as a common feature of the UFD class, a point to which we return in the discussion section. With the exception of Tucana III and Hercules UFDs \citep{Ou:2024}, where long stellar tails undergoing substantial stripping can be seen \citep{Shipp:2018,Li:2018,Erkal:2018} and understood given their unusually small orbital pericenters of only $\sim$3 kpc \citep{Simon:2019,Fritz:2018}. In summary, the current evidence indicates that tidal stripping is uncommon among UFDs, occurring primarily in those with exceptionally small pericenters.

\begin{figure*}
    \centering
 \includegraphics[width=1\textwidth,height=14.5cm]{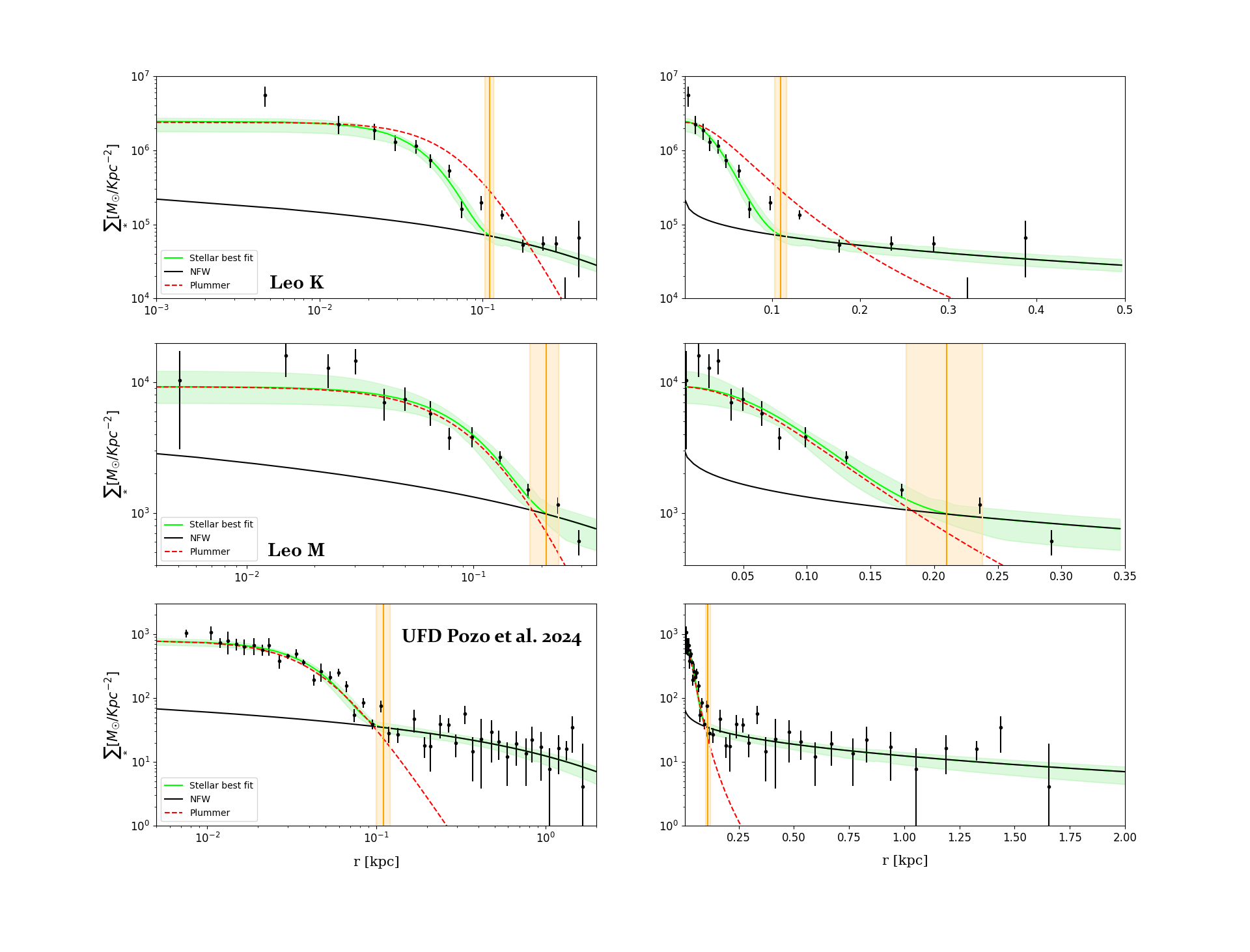}

   \caption{Star-count profiles. The binned star counts for Leo K and M \citep{McQuinn:2024} are compared with the standard Plummer profile shown as the red dashed curve that can be seen to be increasingly in tension at a large radius. The $\psi$DM profile is shown by the green band for our Markov Chain Monte Carlo (MCMC)-based range of acceptable fits, with its soliton core component and an outer Navarro-Frank-White (NFW) profile. The radius of the soliton component is set by the de Broglie scale, and there is a characteristic sharp density drop between the core and the halo predicted in the simulations to be typically a factor of $\simeq 30$, which we mark as the transition radius by the vertical orange band. Note that the right panel has a linear-log scale that helps us appreciate the extent of the outer halo present in the data. In the bottom row, we plot the sum of all star count profiles of recognized UFDs in the Local Group, showing the clear core-halo form and best fitting $\psi$DM profile derived by \cite{Pozo:2024} to illustrate the similarity with Leo K and M.}
    \label{Fig:leos}
\end{figure*}

The recent discovery of the first UFDs beyond the virial radius of the Milky Way at $\simeq 400$~kpc on apparently infalling orbits, named Leo K \& M, provides the first representatives of this class that do not orbit within the halo of the Milky Way \citep{McQuinn:2024}. These three dwarfs are typical of their class in terms of luminosity ($10^4 L_{\odot}$) and size ($R_{1/2} \simeq 30$--80~pc), and therefore clearly point to a simple, undisturbed origin for this class of galaxy, as emphasized by \cite{McQuinn:2024}, rather than transformative tidal losses. The existence of these dwarfs therefore underscores the puzzle of the origin of this distinct galaxy class, with low luminosities ($10^{3-5} L_{\odot}$), low metallicities ([Fe/H] $\simeq -2.5$), small sizes ($R_{1/2} \simeq 50$~pc), and little overlap. Thus, at face value, they appear to demonstrate that UFDs are intrinsically distinct, sitting below the lower end of the general galaxy-mass spectrum, thereby presenting a fundamental puzzle.

Here, we examine the UFD class in the context of wave dark matter ($\psi$DM), since the nearest related class of dSph dwarfs is so well accounted for by $\psi$DM in four important ways. Firstly, we have the presence of a wide core with a density profile that is almost Gaussian \citep{Schive:20142}, which qualitatively matches the characteristic Plummer profiles of the stellar cores of dSph galaxies. Secondly, an outer halo is typically visible, extending well beyond the characteristic $r_c \sim 0.3$~kpc core radius and generally traced to beyond 1~kpc, indicating the presence of a large, low-density spheroidal halo. This is also predicted for $\psi$DM, where the soliton “stands above” the halo in terms of DM density, so that a distinct density transition is predicted at about $2$–$3 r_c$, and this transition has been claimed to be identified in the stellar profiles of deeply detected, well-resolved stellar profiles of dSphs in both the Milky Way and Andromeda \citep{Pozo:2024}. Thirdly, detailed simulations now establish that the spheroidal extended profiles of dSphs can be understood as having been formed by the slow random walk of stars, which is continuously perturbed by the granular density profiles of the halo and the random soliton motion \citep{schive:2020,Dutta:2021,DellaMonica:2023}. Finally, the core DM density is also found to scale with core radius for the dSph class of galaxy, following the steep power-law form predicted by simulations. Since the soliton radius is smaller for a larger mass, $r_{\text{sol}} \propto 1/M_{\text{sol}}$, the central density is higher, which is the opposite behavior to that expected for conventional collisionless gravitating particles, including cold dark matter (CDM), where the DM density naturally increases with galaxy size in a scale-free way. In contrast, for $\psi$DM, the de Broglie scale that sets the core radius is larger for lower mass galaxies, thus predicting opposite behavior.

We recently established that the UFD class of galaxies also appears to follow these same scalings as dSph galaxies, but with a clear offset in the sense that it requires an order of magnitude larger boson mass for the UFDs than for the dSphs \citep{Pozo:2024}. We emphasize that this is not a disaster in the case of $\psi$DM in the generic Axiverse context, where a discrete mass spectrum of axions is generically predicted, spanning many decades in mass depending on the details of dimensional compactification in the string-theory context \citep{Arvanitaki:2010,Dimastrogiovanni:2024,Cicoli:2022}. Thus, in this work, we used new simulations to examine the predicted relations between UFD and dSph galaxies that both appear in terms of the density profiles. In the simulations, we find examples of both light- and heavy-boson-dominated halos, as well as mixtures of the two, with heavy-boson cores within the light-boson-dominated galaxies.

The simulations with two bosonic fields shown in this work are those introduced in \cite{Luu:2025}, where the setup must be carried out with care, since the mutual gravitational interaction couples the fields even though their linear equations evolve independently. The subdominant field, characterized by a lower mean cosmological density, is modulated on all scales by the dominant component, leading to nonlinear evolution with a variety of outcomes. This includes the well-studied nesting of solitons, in which the heavier boson condensate settles within the soliton core of the lighter field, as well as the formation of heavy-boson-dominated halos within the larger filaments shaped by the lighter component. In addition, hybrid halos may emerge, in which the total mass is dominated by the heavy boson but is embedded within an extended, low-density envelope of the lighter field. The resulting galaxy population is therefore not a smooth continuum of properties, it instead depends sensitively on the boson-mass ratio and the relative cosmological abundances of the two fields. Our simulations suggest that the larger the mass difference between the two bosons, the more distinct the resulting galaxy populations, naturally giving rise to bimodality. Moreover, the occurrence of hybrid halos highlights the richness of this two-component scenario, which we expect to explore further with higher resolution simulations.

\begin{figure*}
    \centering
 \includegraphics[width=0.9\textwidth,height=9.5cm]{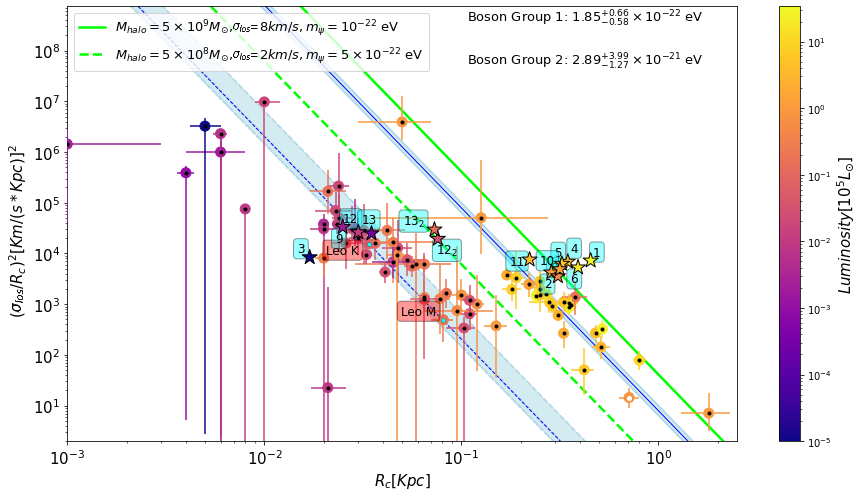}

    \caption{Density versus core radius. The central density is plotted within the fit core radius for each dwarf, estimated as $(\sigma_c / r_c)^2$. Since the values of $\sigma_c$ are not available for the galaxy dataset, we used $\sigma_{los}$ instead. A clear separation between UFD´s and classical dSph´s is apparent with both populations following parallel trends. The slope $d\log \rho_c / d\log r_c = -4$ corresponds to the time-independent soliton solution of the Schrödinger–Poisson equation, where more massive solitons produce narrower cores. The blue lines represent the predicted relation for the UFD´s and dSph´s, whereas the green lines correspond to the input boson masses of the simulation and can be compared with the star symbols (simulated halos) and halo ID number. In the case of halos, $12_2$ and $13_2$ represent halos 12 and 13 in a scenario where they do not experience tidal forces. Note how the dSph galaxies and the non-tidally affected extrapolated UFDs follow their theoretical trend lines. We also mark Leo K ($r_c=0.034^{+0.002}_{-0.019}$Kpc) and Leo M ($r_c=0.081^{+0.01}_{-0.01}$Kpc) galaxies and color-code them by luminosity; this plot provides a prediction of their predicted velocity dispersions, which are as yet unmeasured. Crater II is marked with a white point.}
    \label{Fig:trend}
\end{figure*}

\begin{figure*}

        \centering
        
    \includegraphics[width=0.9\textwidth,height=11.5cm]{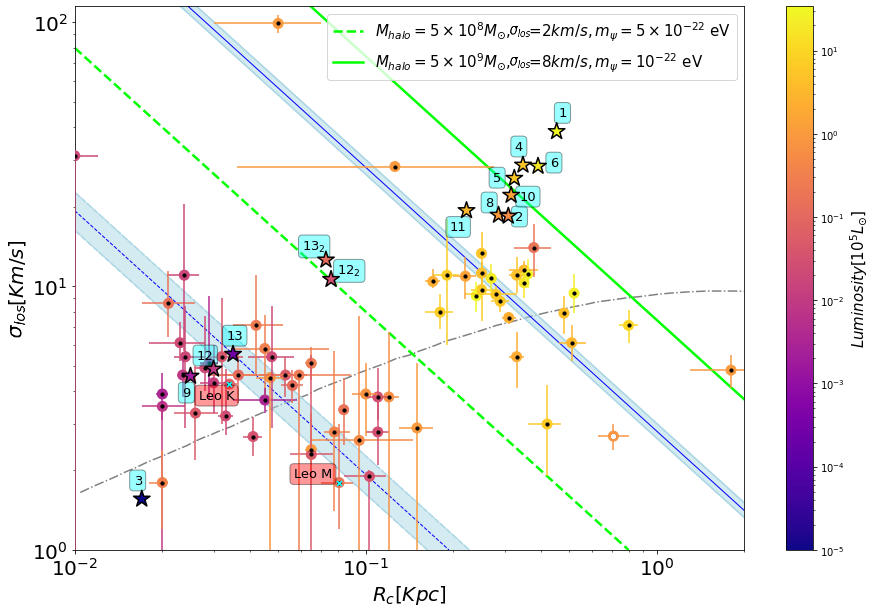}
    \caption{Velocity dispersion versus core radius. The observed velocity dispersion is plotted against the core radius for all dSph and UFD dwarfs, comparing them with the inverse relation, $d\log \sigma_c/d\log r_c=-1$, required by the uncertainty principle; this is shown via diagonal lines. Since the values of $\sigma_c$ are not available for the galaxy dataset, we used $\sigma_{los}$ instead.
 The best fits to the UFD and dSPh dwarfs are indicated in blue, with the corresponding boson masses of $\psi$DM derived from the normalization listed in the legend. The parallel green lines correspond to the boson masses adopted in our simulation, which do not coincide precisely with the best-fit values found in the data (as indicated by the blue lines). Also overlaid is an approximate  CDM-related prediction \citep{Walker:2009} as a dotted gray curve, where galaxies with NFW profiles are naturally predicted to be larger with increasing galaxy mass, i.e., with the opposite sign to the negative-slope $\psi$DM relation. Crater II is marked with a white point.}. \label{figrcd}
 
\end{figure*}

\section{Profile analysis of Leo M and Leo K}

The two newly discovered dwarf galaxies, Leo K and M, were classified by  \cite{McQuinn:2024} as UFDs given their distinct properties in common with the population of this extreme class of dwarf galaxy \citep{Simon:2011,Simon:2015,Simon:2019,Simon:2020}. We first compare the observed stellar density profiles of Leo M and K with the standard Plummer profile, which is traditionally used to describe the stellar profiles, and with the generic soliton profile predicted for $\psi$DM simulations \citep{Schive:2014,Schive:2014b}. We find that the soliton profile matches the cores of all the dSph and UFD galaxy stellar profiles orbiting the Milky Way and Andromeda reported to date well, as shown in our earlier work \citep{Pozo:2024}. In that earlier work, we observed a clear division between an inner cored profile and an outer stellar halo with a distinctive transition radius between the dense inner core and the diffuse outer halo for this class of galaxy \citep{Pozo:2024}. In fitting the stellar density profiles of Leo K and M, we made the simplest assumption that the stars approximately trace the dominant DM density profile, since DM is measured to dominate in all the Milky Way dwarfs, including both the dSph and UFD classes. 

The first simulations in this context revealed a surprisingly rich wave-like structure with a solitonic standing wave core, surrounded by a halo of interference that is fully modulated on the de Broglie scale \citep{Schive:2014}. The solitonic core corresponds to the ground-state solution of the coupled Schr\"oedinger-Poisson equations, with a cored density profile well approximated by \cite{Schive:2014, Schive:20142}:
\begin{equation}\label{eq:sol_density}
\rho_c(r) \sim \frac{1.9~a^{-1}(m_\psi/10^{-23}~{\rm eV})^{-2}(r_{c}/{\rm kpc})^{-4}}{[1+9.1\times10^{-2}(r/r_{c})^2]^8}~M_\odot {\rm pc}^{-3}\,.
\end{equation}
Here, $m_\psi$ is the boson mass, and $r_{c}$ is the solitonic core radius, which simulations show scales as halo mass \citep{Schive:20142} in the following way: 
\begin{equation}\label{eq:rc_c}
    r_{c} \propto m_\psi^{-1}M_{halo}^{-1/3}\,,
\end{equation}

\begin{equation}\label{eq:sol_radius}
r_c=1.6\biggl(\frac{10^{-22}}{m_\psi}  eV \biggr)a^{1/2}
\biggl(\frac{\zeta(z)}{\zeta(0)}\biggr)^{-1/6}
\biggl(\frac{M_h}{10^9M_\odot}\biggr)^{-1/3}kpc
.\end{equation}

Core masses of constant density scale as $\rho_c \propto (\sigma_c/r_c)^2$  (where $\sigma_c$ is the line-of-sight velocity dispersion at the soliton radius), and in the context of $\psi$DM there is also an inverse relationship between soliton core mass and the soliton radius relation required by the nonlinear solution to the Schrodinger-Poisson equation \citep{Schive:2014}; thus, the soliton's density scales more steeply than the volume with radius, i.e., $\rho_c \propto r_c^{-4}$. The radius of the soliton is given approximately by the de Broglie wavelength, $\lambda_{B} =\frac{h}{p}$, following from the uncertainty principle. Here, we use the term uncertainty principle as a convenient analogy. In our classical field simulations, however, the relation simply reflects wave-gradient properties and does not imply any underlying quantum nature. The uncertainty principle is  $\Delta{x}\Delta{p} \geq\frac{\hbar}{2}$, where $\Delta{x}$  is the position dispersion given by the soliton width, 2$\times r_{c}$, and the dispersion in momentum, $\Delta{p}$ is given approximately by $m_{\psi} \sigma$, the product of the boson mass and the velocity dispersion of stars, which are tracer particles of the dominant DM potential. This allowed us to determine the boson mass that corresponds to the de Broglie wavelength: $m_{\psi}\simeq {\hbar}$/$4 r_{c} \sigma_{los}$.\hfill \break

The simulations also show that the soliton core is surrounded by an extended halo of density fluctuations on the de Broglie scale that arise by self interference of the wave function \citep{Schive:2014} that is fully modulated as seen in the simulations \citep{Schive:2014,Mocz:2019} and ``hydrogenic" in form \citep{Hui:2017,Vicens:2018}. These cellular fluctuations are large, with full density modulation on the de Broglie scale \citep{Schive:2014} that modulate the amplitude of the Compton frequency oscillation of the coherent bosonic field, allowing a direct detection via pulsar timing \citep{deMartino:2017,deMartino:2018}.\hfill \break

This extended halo region beyond the central soliton, when azimuthally averaged, is found to follow a Navarro-Frank-White (NFW) density profile \citep{Navarro:1996, Woo:2009,Schive:2014, Schive:20142}, 
so the full radial profile may be approximated as\hfill \break
\begin{equation}\label{eq:dm_density}
\rho_{DM}(r) =
\begin{cases} 
\rho_c(x)  & \text{if \quad}  r< r_{t}; \\\\
\frac{\rho_0}{\frac{r}{r_s}\bigl(1+\frac{r}{r_s}\bigr)^2} & \text{otherwise,}
\end{cases}
\end{equation} 
where $\rho_0$ is chosen such that the inner solitonic profile matches the outer NFW-like profile at approximately $\simeq r_{t}$, and $r_s$ is the scale radius. Setting $\rho_*(r)$ as the stellar density distribution defined by the solitonic $\psi$DM profile:

\begin{equation}\label{eq:stellar_density}
\rho_{*}(r) =
\begin{cases} 
\rho_{1*}(r)  & \text{if \quad}  r< r_t, \\
\frac{\rho_{02*}}{\frac{r}{r_{s*}}\bigl(1+\frac{r}{r_{s*}}\bigr)^2} & \text{otherwise,}
\end{cases}
\end{equation}
where
\begin{equation}\label{eq:stellar_1}
\rho_{1*}(r) = \frac{\rho_{0*}}{[1+9.1\times10^{-2}(r/r_c)^2]^8}~N_* {\rm kpc}^{-3}
.\end{equation}

Here, $r_{s*}$ is the 3D scale radius of the stellar halo corresponding to $\rho_{0*.}$ The central stellar density, $\rho_{02*,}$ is the normalization of $\rho_{0*}$ at the transition radius, and the transition radius, $r_t$, is the point where the soliton structure ends and the halo begins at the juncture of the core and halo profiles. 

We now compare our best-fit soliton+NFW halo profile with the measured stellar profiles of Leo M and K, which are based on the deep star counts reported in \cite{McQuinn:2024} and shown in Figure \ref{Fig:leos}. We see that in both cases a good fit to the joint profile is evident and that for Leo K this joint fit is significantly better than the standard Plummer profile (red curve), which is seen to overshoot the core and to underpredict the shallow outer profile. By comparison with our previous analysis of the full UFD galaxy sample reported to date, we see that Leo K and M are typical of the UFD population and consistent with our previously derived boson mass of $m_{\psi}=2\times 10^{-21}~\mathrm{eV}$, indicating that they are dominated by a significantly heavier boson than the general dwarf galaxy population. We discuss this in a statistical comparison below in the appendix and Figures \ref{Fig:trend} and \ref{figrcd}. Below, we use new simulations of two-field light bosons to show how Leo K and M can be understood in relation to infalling and also tidally stripped halos containing significant proportions of both boson species.

\section{Bimodal distribution in observations and simulations}

We now compare the local observed dwarfs with Leo K and M to examine their claimed UFD joint dynamical and core-profile classification established in \cite{Pozo:2024} for the UFD and dSph dwarfs orbiting both the Milky Way and Andromeda. We first compiled all data published to date in Figure \ref{Fig:trend} for both the UFD´s and dSph dwarfs in the Local Group. This expands on the analysis presented in \cite{Pozo:2024} by incorporating additional newly observed galaxies including the newly reported discoveries of Tucana III, Leo M, Tucana II, Grus II, Boötes I, Leo K, Boötes III, Draco II, Pegasus IV, Aquarius III, Segue I, Centaurus I, Eridanus IV, Segue II, Tucana V, Hercules, Reticulum III, Eridanus III, Delve I, Hydrus II, Cetus II, Leo V, and Triangulum II \citep{Jensen:2024,McQuinn:2024,Cerny:2023,Casey:2025,Foote:2025,Hansen:2024,Ou:2024,Fu:2023,Conn:2018,Simon:2024,Martin:2015,Drlica:2015,Homma:2020,Carlin:2017,Mau:2020}. As can be seen here, the two classes of dwarf are well separated in the plane of observed
core density versus core radius. The density here is derived from the observed velocity dispersion and the core radius derived from the
star counts for each galaxy with a steep relation apparent for both UFD´s and dSph´s between the core radius $r_c$ and the dynamical density estimator predicted for the soliton core density–radius relation, $\rho_c \propto \sigma_{c}^2/r_c^2$. In Figure \ref{Fig:trend}, we color-code by stellar luminosity, revealing two parallel, steeply declining trends: one for UFDs and one for dSphs, with UFDs following a denser and more compact sequence than the dSph´s; this underlines the observed distinction between these two distinct classes of galaxy.

\begin{figure*}
    \centering

 \includegraphics[width=1\textwidth,height=5.5cm]{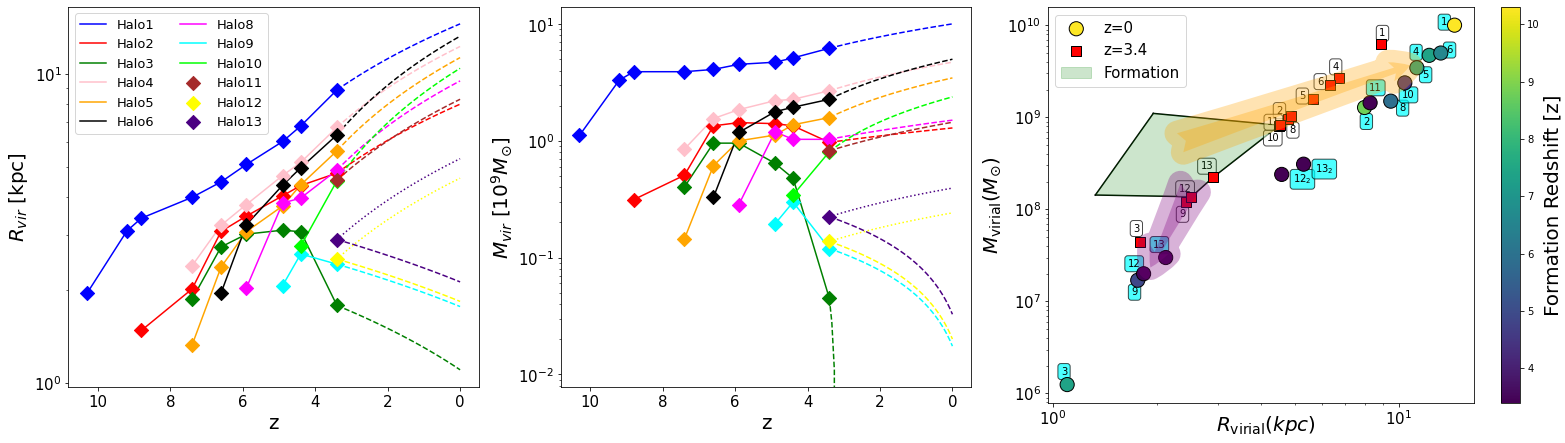}

   \caption{ Evolution of virial mass (left), radius (center), and virial mass versus radius (right). Solid lines trace the evolution of these quantities across all simulation snapshots (as projected in Fig. \ref{Fig:Box}), with extrapolation to $z = 0$ shown as dashed lines, as described in the appendix. Right panel shows virial mass versus radius (left), with the last snapshot marked with red squares (z=3.4), to illustrate that the predicted bimodality is inherent rather than a product of our extrapolations to z = 0.  Arrows in the right panel indicate the evolutionary trend for the two galaxy classes (orange for dSphs and purple for UFDs), with the green area encompassing formation values. The low-mass galaxies all suffer some level of tidal stripping of one or both boson components, and we highlight the sensitivity to tidal effects for the heavy boson-dominated halos 12 and 13 by extrapolating two possible scenarios: one in which they experience tidal stripping similarly to the other UFDs (purple arrow) and another in which they remain unaffected (orange arrow). The latter case is represented by the markers $12_2$ and $13_2$, with the corresponding dotted lines in the figure. All quantities are shown in physical units.    }
   \label{Fig:evolution}
\end{figure*}

The simplest relation required by the uncertainty principle when applied to the soliton is shown in Figure \ref{Fig:trend}. It supports a boson-mass estimate via the scaling $m_{\psi} = \hbar / (2 r_c \sigma_c)$, yielding values that differ by an order of magnitude between UFDs and dSphs. Notably, galaxies associated with both the Milky Way and Andromeda exhibit indistinguishable trends in Figure \ref{Fig:trend}, reinforcing the universality of the two boson populations proposed. This trend explains the observed inverse relationship between core size and velocity dispersion, including outliers such as Crater II and DF44. In contrast to CDM, $\psi$DM predicts smaller velocity dispersions for galaxies with larger cores, as is the case for Crater II. This can be seen in Figure \ref{figrcd}, where the CDM-related prediction \cite{Walker:2009} is represented with a dotted gray line, in opposition to the trends implied by the $\psi$DM relation. This behavior is also consistent with the soliton density--radius scaling relation, $\rho \propto r_c^{-4}$. This larger dataset here continues to show the bimodality established in our previous work where we emphasized the support this provides for the ¨Axiverse¨  interpretation \citep{Luu:2005,Luu:2020,Pozo:2024}.
The bimodality is again observed in Figure \ref{figrcd}, where both galaxy classes are seen to lie in two distinct groups. Moreover, it can be noted that the dSph galaxies tend to have wider core radii and higher velocity dispersions. The solid black line, which represents the predictions for CDM, fails to satisfactorily fit the observed data and simulations.

\begin{figure*}[!ht]
  \sidecaptionvpos{figure}{b}
  \sidecaption
  \begin{minipage}[b]{0.69\textwidth}
    \centering
    \includegraphics[width=.49\linewidth]{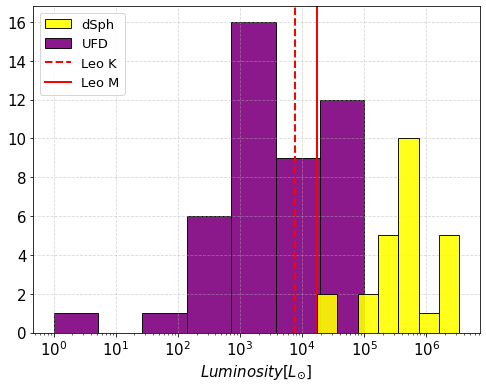}\hfill
    \includegraphics[width=.49\linewidth]{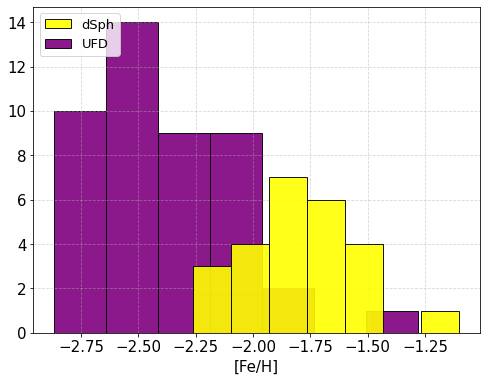}
    \vspace{0.5ex}
    \includegraphics[width=\linewidth]{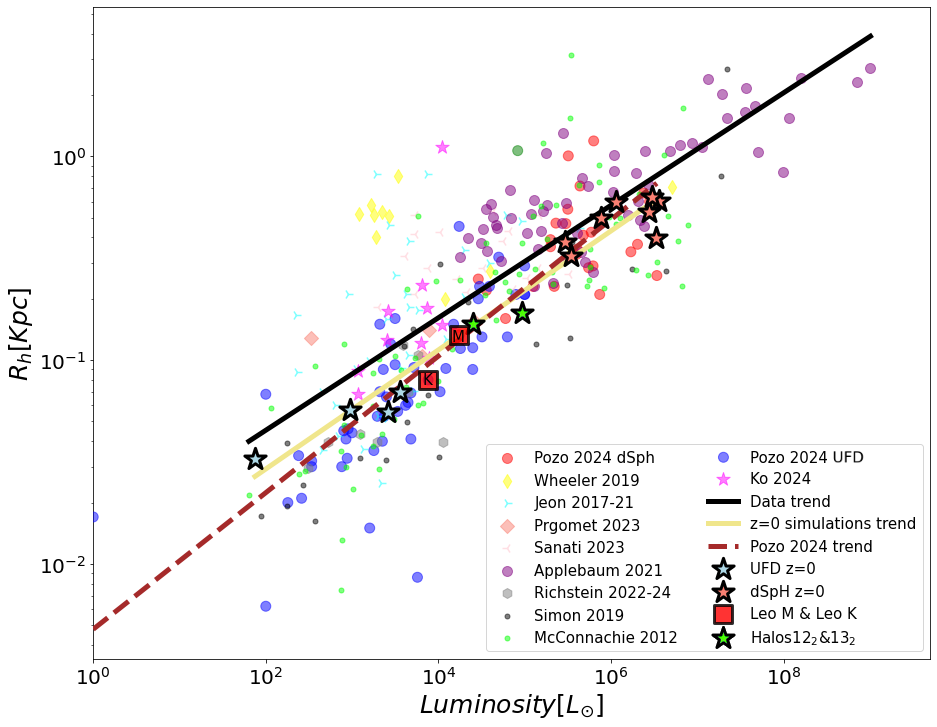}
  \end{minipage}
  \caption{Top left panel: Histogram of luminosities of both galaxy classes. Each galaxy population has been colored differently—yellow for dSphs and purple for UFDs, respectively. We observe that, overall, both classes show a clear distinction in terms of luminosities, forming two separate groups. However, there is no sharp discrete transition between the classes, as some galaxies lie in an overlapping region. Top right panel: Histogram of metallicities of both galaxy classes. We observe the same trend as with the luminosities. Bottom panel: Luminosity versus 2D-projected half-light radius for simulated and observed galaxies. We compare our results (extrapolated to $z=0$ from our latest simulation outputs at $z=3.4$, marked with star-shaped points) with updated catalogs of observed UFD and dwarf galaxies from several works \citep{McConnachie:2012,Simon:2019,Richstein:2022,Richstein:2024}, as well as with simulated galaxies from various studies \citep{Jeon:2017,Jeon:2021,Wheeler:2019,Applebaum:2021,Prgomet:2022,Sanati:2023,Ko:2024}. Additionally, we included the galaxies analyzed in our previous work \citep{Pozo:2024}. For our simulations, we derived the half-light-radius values based on typical proportions corresponding to each dwarf population \citep{Pozo:2024}. We mark Leo K and Leo M with red squares and the locations of halos $12_2$ and $13_2$ with light-green star-shaped points. Note how these two halos still fall within the UFD range, with halo $12_2$ showing similar properties to the recently discovered isolated Leo M UFD (halos $12_2$ and $13_2$ represent halos 12 and 13 in a scenario where they do not experience tidal forces; see Fig. \ref{Fig:evolution}).}
  \label{Fig:comparison}
\end{figure*}

 We also added simulated galaxies, calculated for each galaxy formed in the new two-boson simulation, as described in the appendix.
 The simulated galaxies exhibit density and core-radius values similar to those of the observational data for each population in Figure \ref{Fig:trend}, and they appear slightly shifted to the right corresponding to the boson masses used in the simulations, which are somewhat smaller than those derived from the observed downs and so for this purpose we have added two green lines indicating the expected density–radius ranges for each population given the boson masses used in the simulation (see Figure \ref{Fig:trend}). The simulations show bimodality in a qualitatively similar way to the two observed galaxy populations, where the mass of the different boson progenitor populations is responsible for their distribution. The properties of each of the halos shown above are listed in Table \ref{tabla:collage4} of the appendix.

A limitation of our comparison is that the simulations end at \( z = 3.4 \) due to resolution limitations in solving the Schrödinger–Poisson equations for two fields. However, we extrapolated the growth of these galaxies toward  \( z = 0 \) for an approximate comparison with the data in terms of mass and virial radius by following the evolutionary trend of each halo up to \( z = 3.4 \), as shown in Figure \ref{Fig:evolution}. Since our simulations do not include gas or stellar particles, half-light radii, and luminosities have been estimated using scaling relations observed between these quantities and the virial mass and radius in earlier $\psi$DM simulations, specifically those by \cite{Mocz:2019} and further analyzed in \cite{Pozo:2023}; from there, we were able to extract $r_{vir}-r_{half}$ and $r_{half}-r_{c}$ proportions, respectively. Likewise, core radius values were scaled using the proportionality observed between core radius and half-light radius in our previous study \citep{Pozo:2024}, and velocity-dispersion values were calculated following the virial theorem: $\sigma_{los} = ({G M_{vir}/{3R_{vir}}})^{0.5}$. In Figure \ref {Fig:comparison}, we compare the properties of our simulated galaxies with various datasets. Specifically, we contrasted our results with updated catalogs of observed UFD and dwarf galaxies from several works \citep{McConnachie:2012,Simon:2019,Richstein:2022,Richstein:2024}, as well as with simulated galaxies from different studies \citep{Jeon:2017,Jeon:2021,Wheeler:2019,Applebaum:2021,Prgomet:2022,Sanati:2023,Ko:2024} and with the sample of galaxies analyzed in Pozo et al. (2024). In the bottom panel, our simulated galaxies fall within the observational range, showing good agreement in luminosities and half-light radii for both populations. Additionally, we plot trend lines corresponding to the different data sources, and notably, the trend line from our simulations aligns almost perfectly with one of them (black line; product of previous catalogues \citep{McConnachie:2012,Simon:2019,Richstein:2022,Richstein:2024} and simulations \citep{Jeon:2017,Jeon:2021,Wheeler:2019,Applebaum:2021,Prgomet:2022,Sanati:2023,Ko:2024}), while remaining consistent with the others. We can see that overall the UFD and the dSph are also divided in terms of luminosity and size. UFDs are generally smaller and less luminous, and, as observed in the lower panels, there is a clear distinction between both populations in terms of metallicity, with UFDs tending to be metal poorer. It is also true that Figure \ref{Fig:comparison}, while still presenting a distinction between the two groups, reveals that the transition is not abrupt, as some of the brightest UFDs are similar to the faintest dSphs. In summary, regardless of luminosity, we can distinguish UFDs from dSphs by their other properties. However, once we have done that, we see that the luminosity value is a good indicator of whether a galaxy will be classified as UFD or dSph based on other properties (in particular, metallicity, radius, or density if sigma is measured). Of course, if we only know the luminosity, we would not know that UFDs and dSphs are distinctly different classes, even if the region where they overlap is pretty small, as can be seen in the histograms in the upper panels of Figure \ref{Fig:comparison}. The non-universal classification of what constitutes a UFD also has an impact on the number of galaxies that lie in this overlapping region.

Figure \ref{Fig:Box} shows the evolution of the full simulation box for both the separated and combined fields. This figure illustrates the expected scenario in which UFDs are formed within filaments near classical dSphs well. This behavior is clearly seen in the four highlighted UFDs: galaxy 3 forms within a filament connected to classical dwarf 1, and, similarly, galaxy 9 forms in the filament of galaxy 4 at redshift $z = 4.9$. A comparable situation is observed for galaxies 12 and 13, which are born within the filaments of galaxies 1 and 8, respectively, at $z = 3.4$.

\section{Tidal effects}

In the context of $\psi$DM, tidal effects are known from simulations to readily strip the low-density outer halo, whereas the soliton is more robust, as the ground state is self-reinforcing \citep{Chan:2025}. It can be seen in Figure \ref{Fig:leos} that the soliton is visibly prominent compared to the surrounding stellar halo usually found extended around dSph galaxies, and similar behavior is also visible for UFD´s (as we show in the lower panel of Fig. 1) on a smaller radial scale. We have argued that tidal stripping can account for the prominence of the observed cores in particular for dSphs on small orbits well within the Milky Way. This quantitatively supports the level of halo stripping predicted for dwarfs orbiting the Milky Way by \cite{schive:2020}, which was further quantified by  \cite{Pozo:2022}. In the case of Crater II, we show that its origin is likely a standard dSph that has been strongly stripped, such that its low velocity dispersion and wide core radius follow from the core-halo relation \citep{Pozo:2022}. This tidal stripping interpretation is now reinforced by the long tails reported in GAIA-based astrometry \citep{Fritz:2018}, which have established both tidal tails and a close pericenter orbit for Crater II, clearly demonstrating the transformative role of tidal stripping in this case. In Figures \ref{Fig:trend} and \ref{figrcd}, we can see that Crater II, marked with a dark blue dot, continues to lie on the predicted relations for the soliton core (see Figure 1 of \cite{Pozo:2024} for more details), despite the stripping of its outer halo. We stress that this is consistent with the expected expansion of the core radius as the halo mass is reduced.

 In our previous predictions \citep{Pozo:2024}, we suggested that galaxies could in principle lie anywhere along predicted scaling relations, although the data are relatively concentrated to the dSphs typically lying below the UFDs in density,  a feature that the new simulations suggest can be explained by $\psi$DM dynamics (see Figure \ref{Fig:trend}). In this context, we expect that although most dwarfs in each class should cluster near the upper end of the two distributions, since more galaxies are born near the lower mass limit given the rising slope of the mass function, there will be a tidal spread extending toward lower mass, lower density, and lower velocity dispersion; it will be wider in terms of core radius. Halo 3 is a clear example of a stripped system, with a wide, low-density halo, similar to systems such as Hercules and Tucana III. While most do not show clear tidal imprints, some UFDs are known to have tidal tails \citep{Ou:2024,Li:2018}. Additionally, we observe that the tidal forces typically experienced by some UFDs may significantly affect their evolution, causing them to deviate leftward from their theoretically predicted position. This effect is illustrated by the different locations of halos 12 and 13, depending on whether or not tidal interactions are taken into account in their evolutionary history. This may also explain why, in Figure \ref{Fig:trend}, we find a greater number of UFD galaxies located to the left of the expected trend (blue line) compared to dSphs, as UFDs are more susceptible to tidal disruption. However, these tidal forces are not the origin of the two distinct galaxy populations, as shown in Figures  \ref{Fig:trend}, \ref{figrcd}, and \ref{Fig:evolution}.

In our new two-boson simulations, we see that tidal effects are also important, particularly for the low-mass galaxies, where in two cases (halos 3 and 9) they were transformed by stripping, significantly reducing their masses. We also have two additional heavy-boson-dominated low-mass halos (12 and 13) that have not yet evolved long enough in the simulation to determine whether they will eventually experience tidal disruption. Therefore, further evolution will be needed to assess their susceptibility to tidal forces, and the overall picture remains mixed. We emphasize that the simulations presented here are not yet close to the larger boson mass ratio implied between the UFD and dSph populations in Figures \ref{Fig:comparison}, \ref{Fig:trend}, \ref{figrcd}, and 4. Hence, we can only make qualitative statements at this point, until computational improvements allow for a sufficiently large dynamical range to explore the two-boson case more fully.

The limitations of the parameter space explored by our current simulations prevent us from making definitive claims regarding the effects of tidal stripping or the precise DM composition in these galaxies. In particular, the simulations likely overestimate the impact of tidal stripping, since a heavier boson mass, as suggested by observations (see Figure 3), would result in denser UFDs than those produced here, making them more resistant to tidal disruption. Indeed, if the actual boson mass ratio is closer to 20, as observational data indicate, the heavy field would be significantly denser than in our current model. This increased density would allow the UFDs to better withstand tidal forces, thereby reducing the role that tidal stripping plays in their evolution. Consequently, the influence of tides on UFDs in our simulations should be viewed as an upper limit, rather than a realistic outcome.

\section{Conclusions}

We show that in terms of their stellar profiles, Leo K and M are both consistent with the population of UFD´s by comparison with all other Local Group dwarfs published to date in this galaxy class, as we show in Figure \ref{Fig:trend}. Despite their low luminosities, the star counts of Leo M and Leo K have recognizable stellar cores that resemble a Plummer profile, but with a closer match to the soliton solution of $\psi$DM in its ground state, as shown in Figure \ref{Fig:leos}. The soliton fit has no adjustable parameters beyond the boson mass, which uniquely determines the soliton radius. In both of these dwarfs, the cores appear surrounded by an extended stellar halo that extends to over 1kpc, as generically predicted for $\psi$DM for the excited states that surround the solitonic core, approximately following an NFW profile when averaged azimuthally, as generically predicted by $\psi$DM simulations \citep{Schive:2014}, reflecting the non-relativistic (i.e., cold) definition of a Schrodinger-Poisson condensate.

We demonstrate that a two-field $\psi$DM scenario with two distinct boson-mass populations can successfully reproduce both dark-matter-dominated classes of dwarf galaxies, namely the dSph and UFDs and the trend observed in the data. Moreover, the simulation-generated galaxies that exhibit the observed distinct differences in mass, size, and luminosity observed between these classes of dwarf galaxies now well established from many detailed deep imaging, spectroscopy, and astrometric observations in the Local Group.

It is important to note that, due to the limitations inherent to $\psi$DM simulations \cite{Luu:2025}, the boson mass of the heavy field in this is not as large as the value suggested by observational data \citep{Pozo:2024}, for which a value 20 times larger than the light-boson mass indicated by the data, rather than the factor-of-five mass ratio of our simulations. Additionally, the relative density ratio between the two boson populations should favor the light field over the heavy one (compared to the setup in this simulation), highlighting the need for future simulations with significantly higher spatial and temporal resolution in order to match with the empirically inferred boson masses, and if possible to extend such simulations to lower redshifts approaching \( z \sim 0 \), which is currently out of reach given the computational limitations. We also emphasize that a more massive boson for the heavy field would make UFDs more robust against tidal forces \citep{Chan:2025}, leading to a higher proportion of UFD-type dwarfs that evolve to the present day, such as Leo K and Leo M \citep{McQuinn:2024}. Remarkably, even with the approximate current simulation setup, our most general finding of a clear bimodality in terms of the formation of heavy-boson-dominated dwarfs together with larger light-boson-dominated and mixed examples does support the core idea that the Axiverse framework \citep{Arvanitaki:2010} can offer a plausible, uncontrived  explanation for the existence of the puzzlingly distinct dwarf-galaxy populations now empirically established. 

As there have not been simulations for two bosons with baryon physics included, it is too early to say whether galaxy distribution will follow that of DM. However, we have evidence that DM properties are not continuous when varying the ratio of two bosons, as shown in \cite{Luu:2025}. Specifically, when their density ratio is below a certain threshold, soliton cores of the heavy boson cannot form due to gravitational heating of light-field granules, which breaks the bimodality. This threshold in turn depends on the boson mass ratio. Thus, we found the bimodality of two boson cases is only viable at a certain density ratio range (see discussion in \cite{Luu:2025}).

These UFDs are expected to be some of the oldest galaxies, especially when compared to the less metal-poor classical dwarf galaxies. In general, it is understood that their stellar formation occurred earlier and over a shorter period than in classical dwarfs, with all their stars being formed at \( z > 6 \), and in some extreme cases as early as \( z \simeq 10 \) \citep{Weisz:2014,Brown:2014}. In the $\psi$DM framework, the suppression of small-scale structure formation arises from the quantum nature of ultra-light bosonic DM, where bosons of \( m_\psi \sim 10^{-22} \,\mathrm{eV} \) correspond to large de~Broglie wavelengths, so the cutoff in the matter power spectrum occurs at a smaller scale. On the other hand, a lower boson mass generates suppression and a larger cutoff scale. Therefore, in a $\psi$DM scenario, UFDs that formed earlier than classical dwarfs are naively interpreted as the result of gravitational collapse in regions dominated by higher mass bosons and potentially explaining why UFDs appear older and metal poorer than their classical counterparts. UFDs tend to form in high-density regions, particularly in filaments where the heavy field dominates more strongly, raising the question of whether UFDs are necessarily older than classical dwarfs, as explored in \citet{Pozo:2023}. Addressing whether this also holds true for their stellar populations will require new simulations including baryonic physics and stellar particles,  and we should keep in mind that the well-known age–metallicity degeneracy, whereby a younger metal-poor isochrone is similar to older, metal-richer stars with differences too subtle to distinguish observationally, especially in systems older than 8~Gyr. This is consistent with what we observe in the simulations: galaxies that are dominated by the heavy field at birth tend to form preferentially within these dense, filamentary regions. It suggests that UFDs could indeed form early, but likely in smaller numbers than previously expected due to the limited capacity of the heavy field to form structures without assistance from the presence of the lighter boson field that takes longer to collapse. More detailed simulations, especially those incorporating baryonic physics, allowed us to address whether mixed-field dynamics dominates in high-density environments and the extent to which an increased light-boson population affects the formation redshift of heavy-boson-dominated dwarfs.

Finally, we emphasize that the two-boson solution proposed here for the UFD and dSph dwarf galaxies may be definitively tested using pulsar timing, as two monotone timing residuals will be imprinted corresponding to the respective oscillating scalar fields, each at precisely twice the Compton frequency: $$2m_\psi c^2/h=0.66yrs^{-1} (m_\psi/10^{-22}eV),$$ with an amplitude of 
$$\delta t=100ns(m_{\psi}/10^{-22}eV)^3 (\rho_\psi/10GeV^{-3})$$ and a period of about four months.
This signal is strongest for pulsars near the Galactic center. There, the peak DM mass density is nearly two orders of magnitude larger than the local 
value at the solar radius of $0.4GeV^{-3}$ in the case of the soliton predicted for the canonical light boson ($m_\psi=10^{-22}$eV) and within the range of SKA sensitivity within the soliton radius of $r_{sol}=80pc(m_\psi/10^{-22}eV)$ for the Milky Way \citep{deMartino:2017,Luupulsar:2024,Boddy:2025}. For the heavier boson of $10^{-21}$eV, the reduced detectability by $m_\psi^{-3}$ at a fixed mass density is countered by $m_\psi^2$ from the correspondingly denser soliton; so, the predicted amplitude of the timing residual is $\delta t=10ns$ over a frequency of six days within the soliton radius of 5pc. Conceivably, this signal may also be detectable for suitable millisecond pulsars within the central nuclear star cluster of the Milky Way.

\begin{acknowledgements} 
We warmly acknowledge Douglas Finkbeiner for very fruitful conversations. A.P. is grateful for the continued support of the DIPC postgraduate program as well as the Center for Astrophysics | Harvard \& Smithsonian for their warm hospitality at the beginning of this project. R.E. acknowledges the support from grant numbers 21-atp21-0077, NSF AST-1816420, and HST-GO-16173.001-A as well as the Institute for Theory and Computation at the Center for Astrophysics. We are grateful to the supercomputer facility at Harvard University where most of the simulation work was done. G.S. is grateful to the IAS at HKUST for their generous support. TJB supported by the Spanish grant PID2023-149016NBI00 (funded by MCIN/AEI/10.13039/501100011033 andby “ERDF A way of making Europe”. P.M. acknowledges this work was in part performed under the auspices of the U.S. Department of Energy by Lawrence Livermore National Laboratory under contract DE-AC52-07NA27344, Lawrence Livermore National Security, LLC.

\end{acknowledgements} 

\section*{Data availability}

The data underlying this article will be shared on reasonable
request to the corresponding author.

\bibliographystyle{aa} 
\bibliography{apssamp}

\newpage
\begin{appendix}

\onecolumn

\section{Supplement: simulations of 2-boson $\psi$DM} \label{secsimu} 

 We summarise our simulations of two coupled fields of light bosons (2$\psi$DM) with which we can go on to interpret the UFD galaxies Leo K and M, and the UFD and dSph classes of dwarfs in general. In the non-relativistic limit, the dynamics 2$\psi$DM is governed by coupled Schrödinger–Poisson equations for two fields, $\psi_1$ (light, $m_1$) and $\psi_2$ (heavy, $m_2$), evolving under a shared gravitational potential $\Phi$. These are solved numerically using a fully parallelised pseudo-spectral method, as described fully in \cite{Nhan:2023}. A high-resolution cosmological simulation has been performed with $m_1 = 10^{-22}$eV, $m_2 = 5 \times 10^{-22}$eV, and a density ratio $\beta_2 \equiv \Omega_2/\Omega_m = 0.7$, within a $1.7~{\rm Mpc}/h$ box at $2048^3$ resolution, and although the chosen masses may be in some tension with observations, in any case the scale-invariant nature of the equations ensures their physical relevance. The limitations in the boson mass range, together with a density ratio of $\beta_2 = 0.7$---which results in the heavy boson population being dominant (opposite to what we expect)---are imposed by current constraints in $\psi$DM simulations. While these limitations affect the reliability of certain galaxy properties, such as the detailed composition of each boson component within galaxy profiles, the two-boson mass model remains a useful and insightful framework to test. In particular, it allows us to investigate whether it can account for the observed bimodality between UFD and dSph galaxy populations. 

 Initial conditions are generated via linear perturbations and a modified version of the Code for Anisotropies in the Microwave Background (\texttt{CAMB}), with wave functions initialized via the Madelung transform equations. The simulation evolves from $z = 127$ to $z = 3.4$, with minor under-resolution at the smallest scales that does not affect the main results \cite{Nhan:2023} for more details. The simulation are not evolved below z=3.4, given the limitations of increasing box size and so in making a comparison with the locally predicted galaxies at the present day some extrapolation is needed, which although approximate in terms of mass and radius serves as a useful guide as we show below for most galaxies where growth in mass is smoothly convergent.

 \begin{figure*}[htbp]
   \centering
 \includegraphics[width=0.49\textwidth,height=8.5cm]{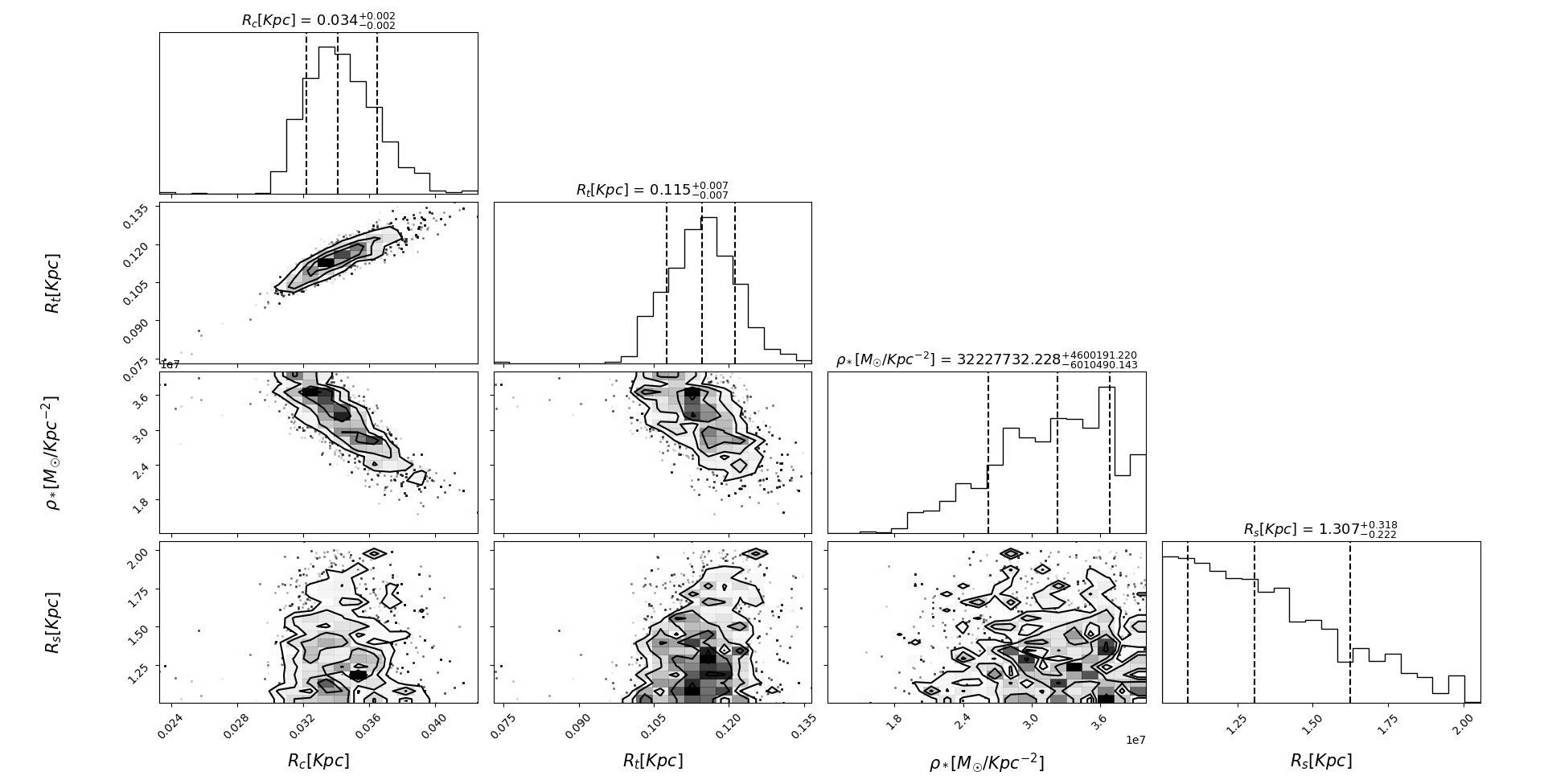}
\includegraphics[width=0.49\textwidth,height=8.5cm]{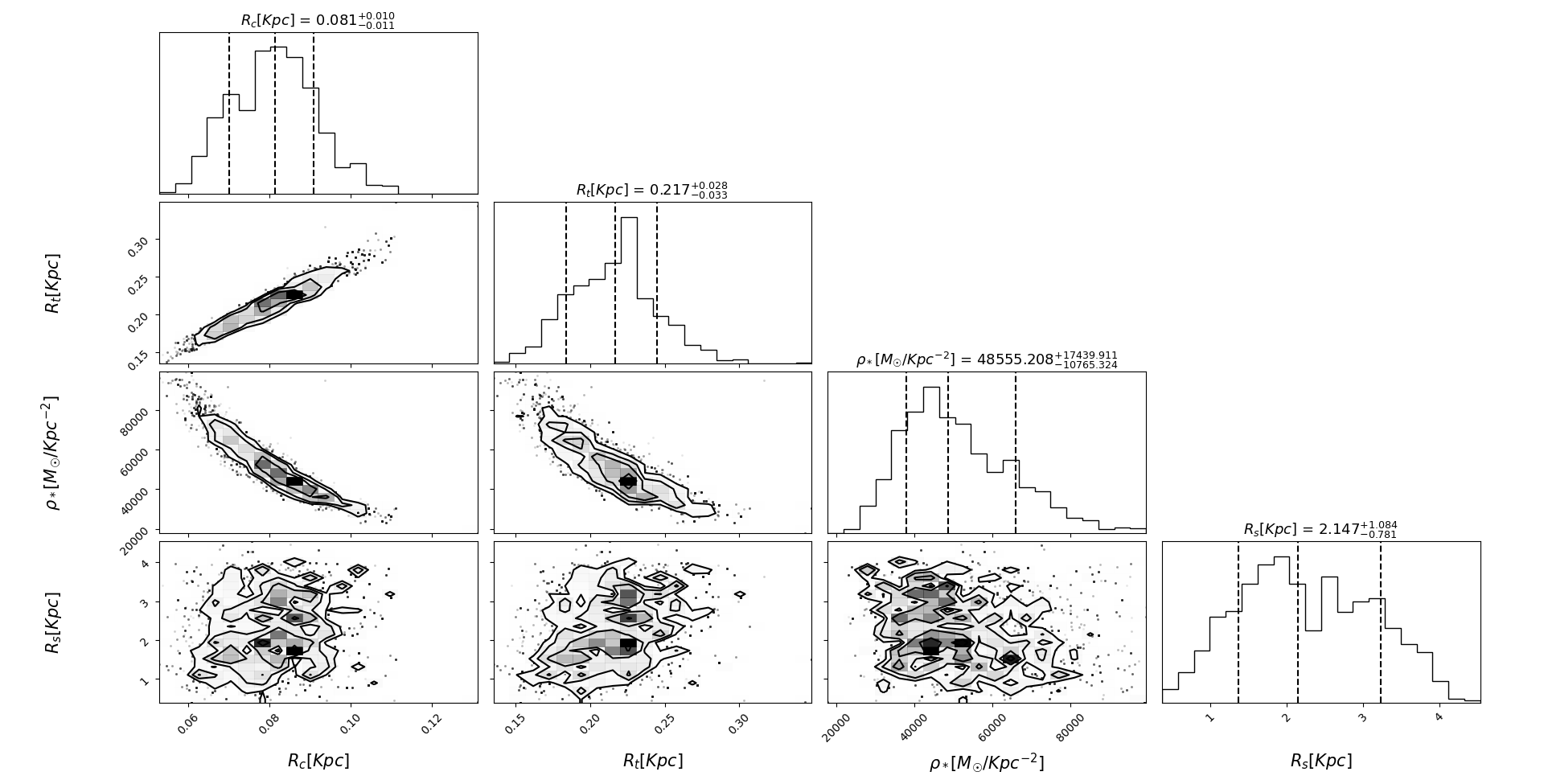}

  \caption{All dSph: Classical dwarfs mean profile (Figure \ref{Fig:leos} top panel): correlated distributions of the free parameters.  As can be seen the core radius and transition radius are well defined despite wide Gaussian priors, indicating a reliable result. The contours represent the 68\%, 95\%, and 99\% confidence levels. The best-fit parameter values are the medians (with errors), represented by the dashed black ones.}
   \label{Fig:corner}
\end{figure*}

\begin{table*}[htbp]
\caption{Summary of the simulated galaxies at the final redshift of the simulation (z = 3.4)}              
\label{table:1}      
\centering                                     
\begin{tabular}{|c|c|c|c|c|c|c|c|c|c|}
\hline
Halo ID& $M_{vir}$ & $R_{vir}$ & $R_{half}$& $R_{c}$ & $\sigma_{v}$   & $L_{*}$ & $z_f$ & $\psi_1/\psi_2$ & Class \\
& $10^9 M_{\odot}$&kpc&kpc&kpc&km/s &$10^5 \times L_{\odot}$&&& \\
\hline
${1}$ &8.40 & 43.19 &$0.42^{+0.60}_{-0.24}$&$0.31^{+0.45}_{-0.19}$&8.29 &$4.99^{+8.90}_{-1.09}$ &10.3 & 0.90 & dSph \\
\hline
${2}$ &1.23 & 22.77 &$0.23^{+0.33}_{-0.13}$ &$0.17^{+0.25}_{-0.10}$ &  4.45 &$0.78^{+1.40}_{-0.2}$&8.8 & 1.04& dSph \\
\hline
${3}$  &0.10 &9.99&$0.06^{+0.09}_{-0.02}$&$0.03^{+0.05}_{-0.02}$& 1.64 & $0.008^{+0.01}_{-0.004}$& 7.4& 7.05 &UFD \\
\hline
${4}$  &3.29 &31.61&$0.32^{+0.46}_{-0.18}$&$0.24^{+0.35}_{-0.14}$&  6.24&$2.14^{+3.81}_{-0.47}$ &7.4 & 0.59 &  dSph \\
\hline
${5}$  &1.86 &26.14&$0.27^{+0.38}_{-0.15}$&$0.20^{+0.29}_{-0.12}$& 5.20 &$1.25^{+2.22}_{-0.27}$ &7.4 & 0.39 & dSph\\
\hline
${6}$ &3.06 &30.85&$0.30^{+0.43}_{-0.17}$&$0.22^{+0.32}_{-0.13}$& 6.00 &$1.87^{+3.34}_{-0.41}$ &6.6 & 0.70 & dSph  \\
\hline
${8}$ &1.22 &22.77&$0.23^{+0.33}_{-0.13}$&$0.17^{+0.25}_{-0.10}$&  4.55 &$0.83^{+1.48}_{-0.18}$ &5.9 & 0.42 &dSph  \\
\hline
${9}$   &0.16 &11.65&$0.08^{+0.13}_{-0.03}$&$0.04^{+0.07}_{-0.02}$& 2.22 &$0.02^{+0.04}_{-0.0001}$ &4.9 & 1.74 & UFD \\
\hline
${10}$ &1.04 &21.59&$0.22^{+0.31}_{-0.12}$&$0.16^{+0.23}_{-0.09}$&   4.20&$0.65^{+1.16}_{-0.14}$ &4.4 & 0.39 & dSph \\

\hline
${11}$&0.97 &21.03&$0.22^{+0.31}_{-0.12}$&$0.16^{+0.23}_{-0.09}$&  4.22 &$0.66^{+1.18}_{-0.14}$ &3.4 & 0.45& dSph \\

\hline
${12}$ &0.16 &11.53&$0.08^{+0.13}_{-0.03}$&$0.04^{+0.07}_{-0.02}$&  2.28 &$0.02^{+0.04}_{-0.0001}$&3.4  & 0.005 & UFD \\

\hline
${13}$ &0.25 &13.48&$0.10^{+0.15}_{-0.04}$&$0.05^{+0.08}_{-0.03}$&  2.67 &$0.03^{+0.06}_{-0.0002}$ &3.4  & 0.13 & UFD \\
\hline
\hline
\hline
\end{tabular}
\tablefoot{Column 1: Halo ID, Column 2: Halo mass,  Column 3: Virial Radius, Column 4: Extrapolated half-light radius, Column 5: Extrapolated core radius, Column 6: Computes velocity dispersions, Column 7: Inferred luminosities, Column 8: Formation redshift, Column 9: mass ratio of each boson species within $R_{vir}$, Column 10: Assigned class.} 
\label{tabla:collage4}
\end{table*}

\begin{figure*}
   \centering
\includegraphics[width=0.4\textwidth,height=4.75cm]{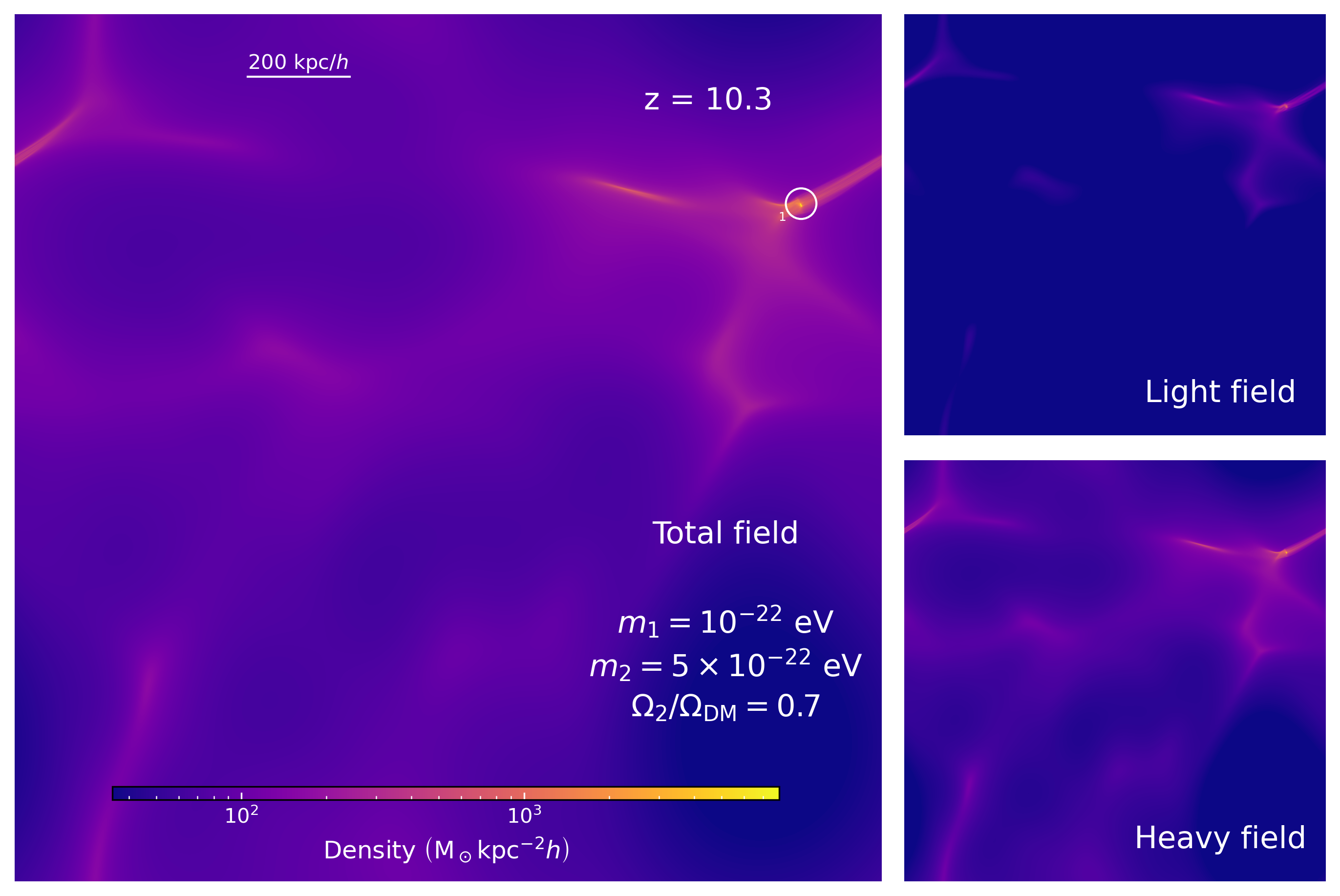}
 \includegraphics[width=0.4\textwidth,height=4.75cm]{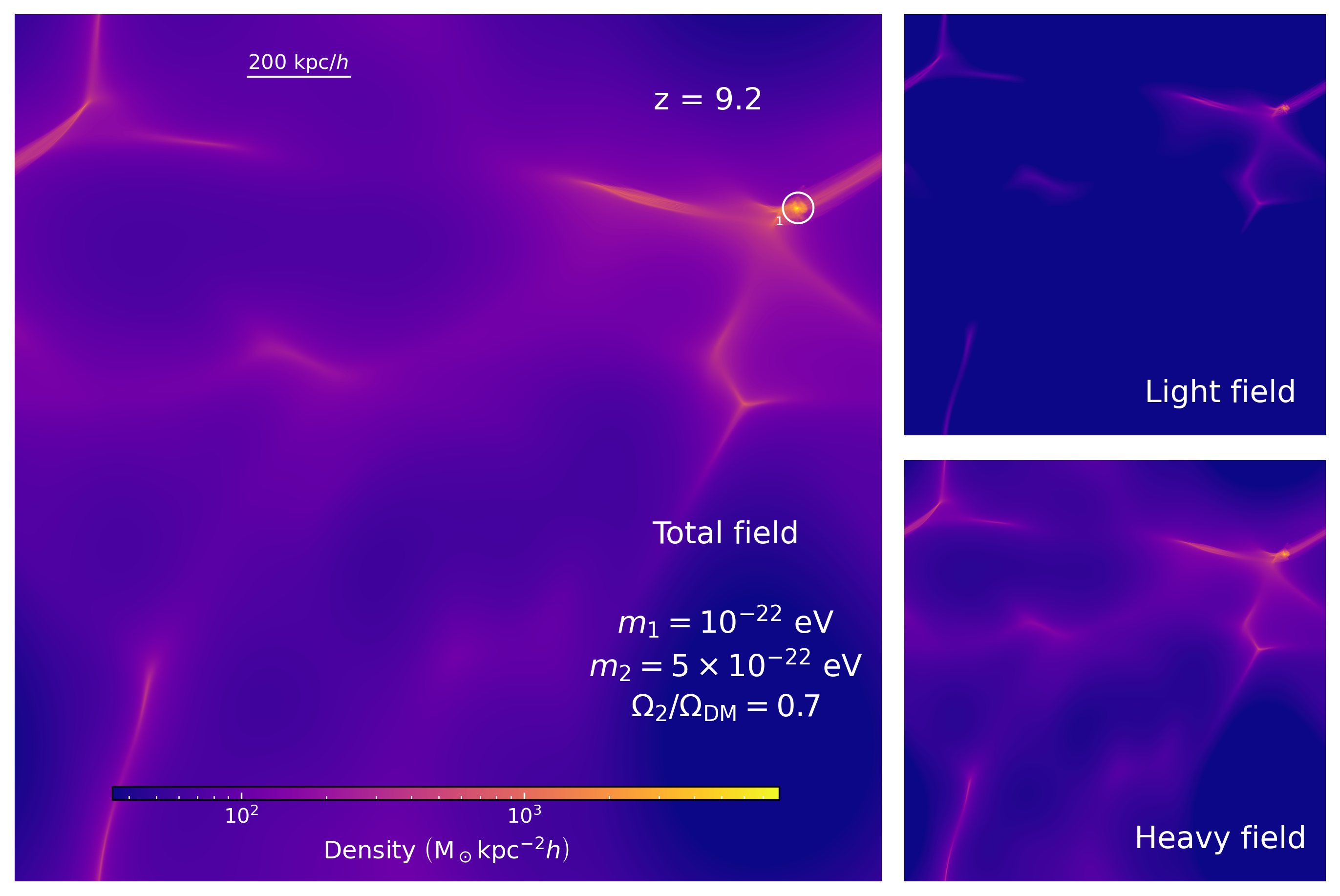}
\includegraphics[width=0.4\textwidth,height=4.75cm]{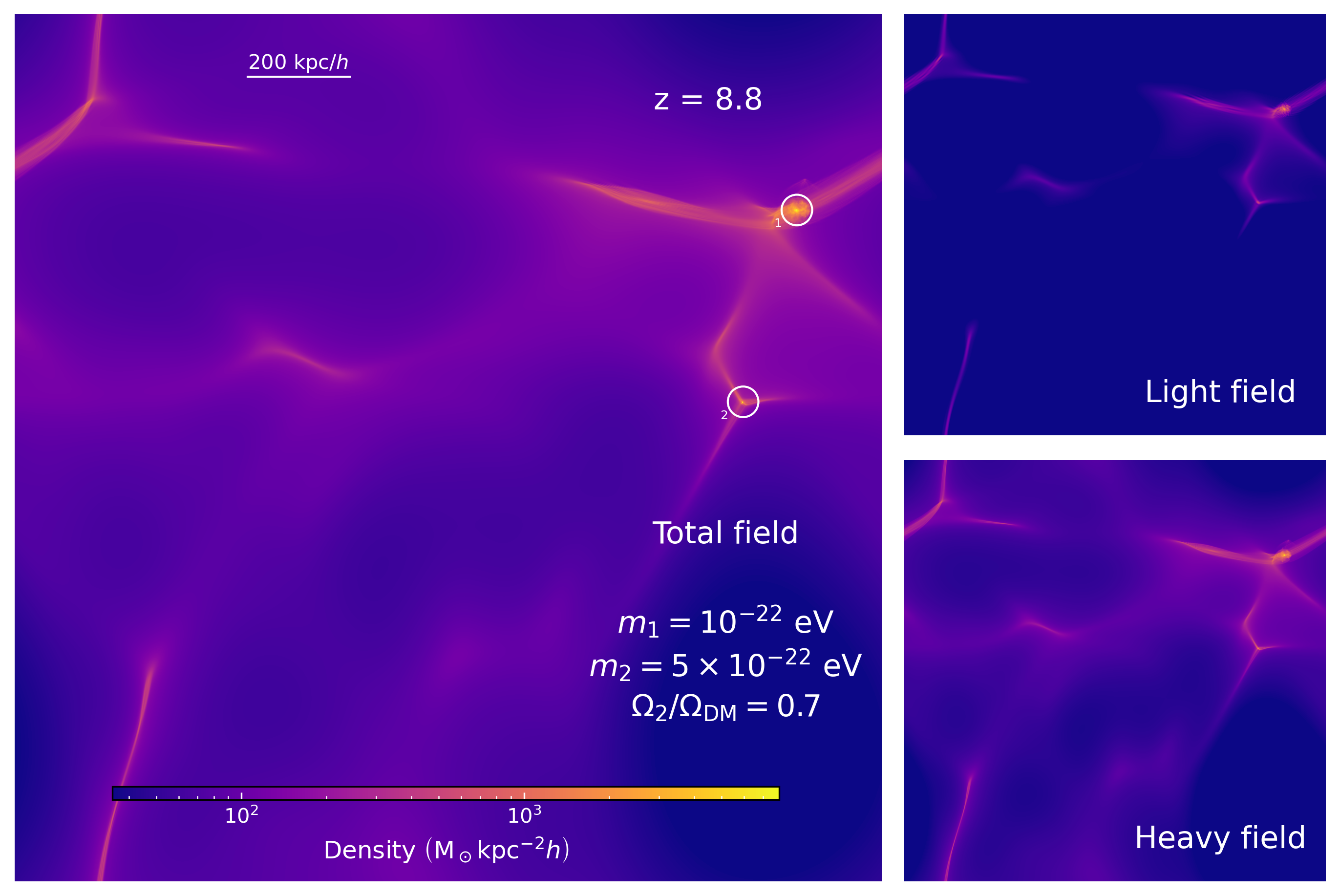}
\includegraphics[width=0.4\textwidth,height=4.75cm]{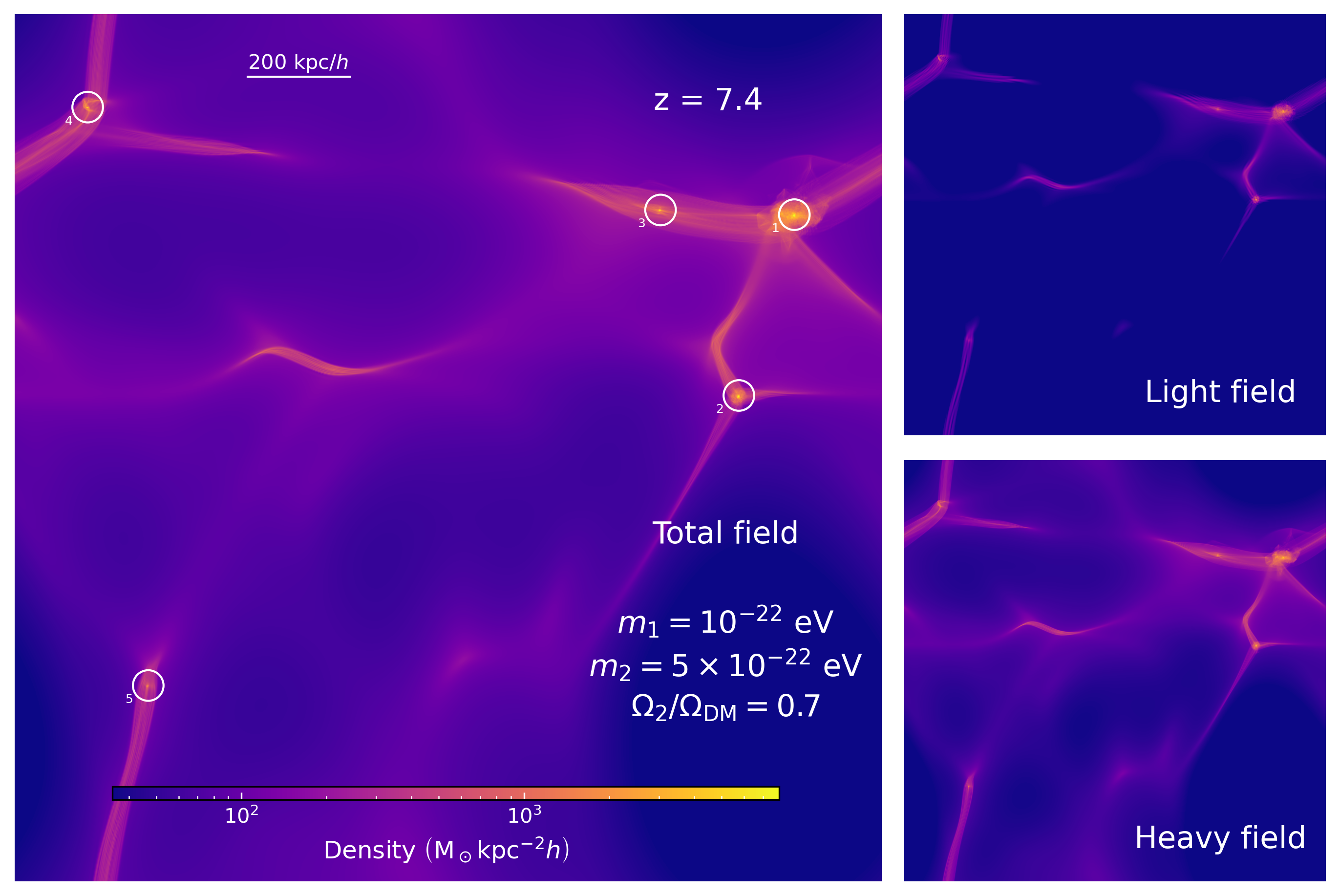}
 \includegraphics[width=0.4\textwidth,height=4.75cm]{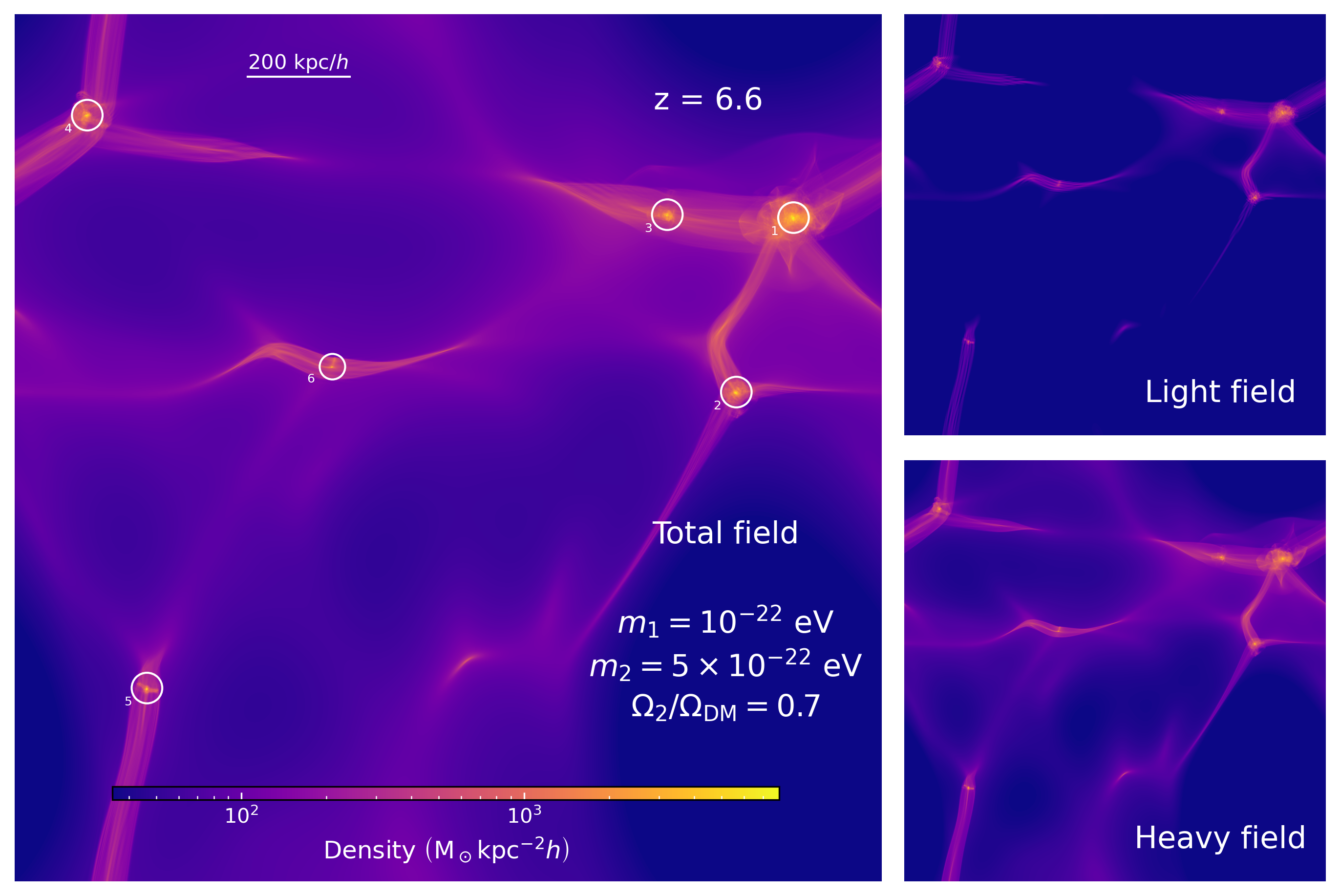}
\includegraphics[width=0.4\textwidth,height=4.75cm]{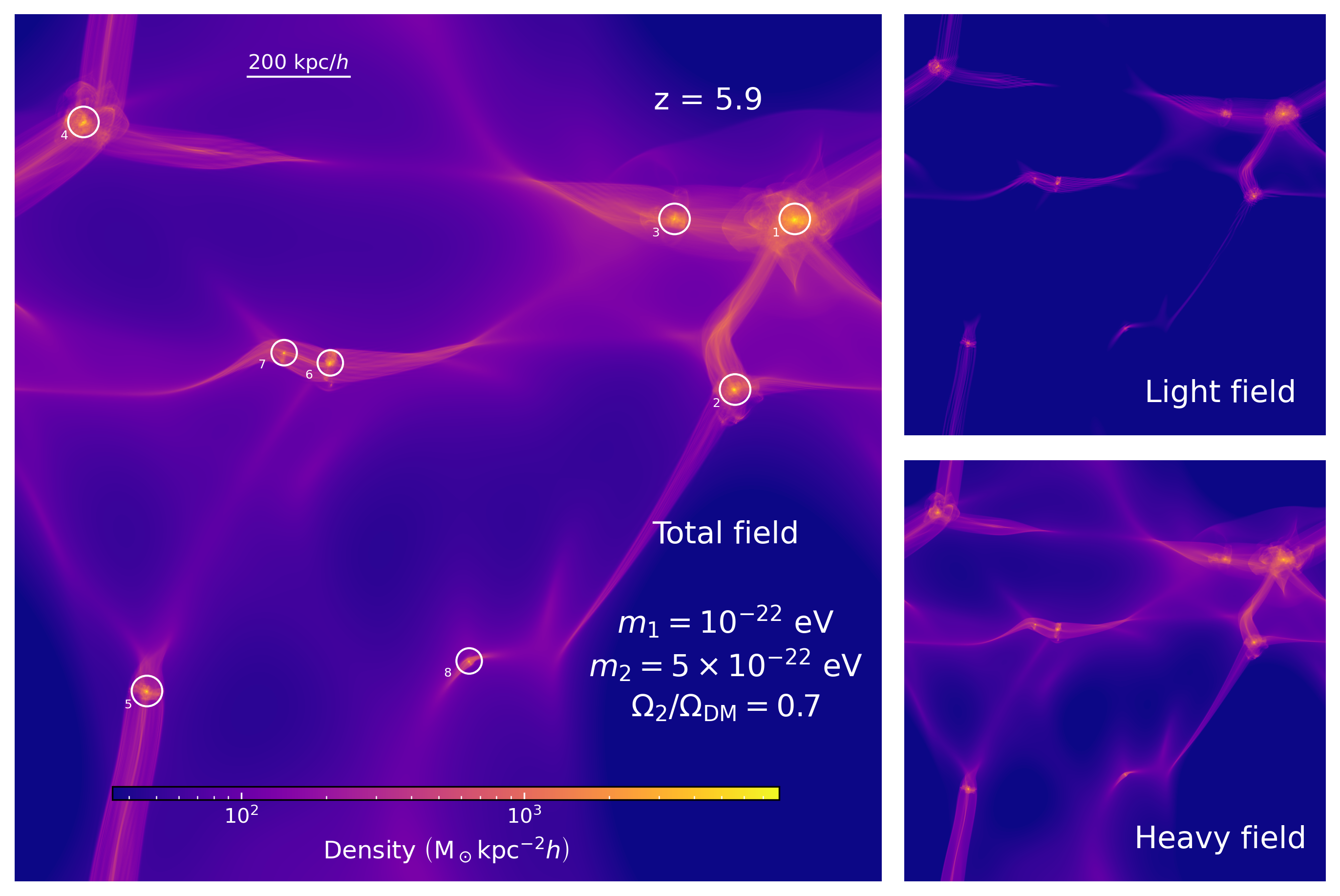}
  \includegraphics[width=0.4\textwidth,height=4.75cm]{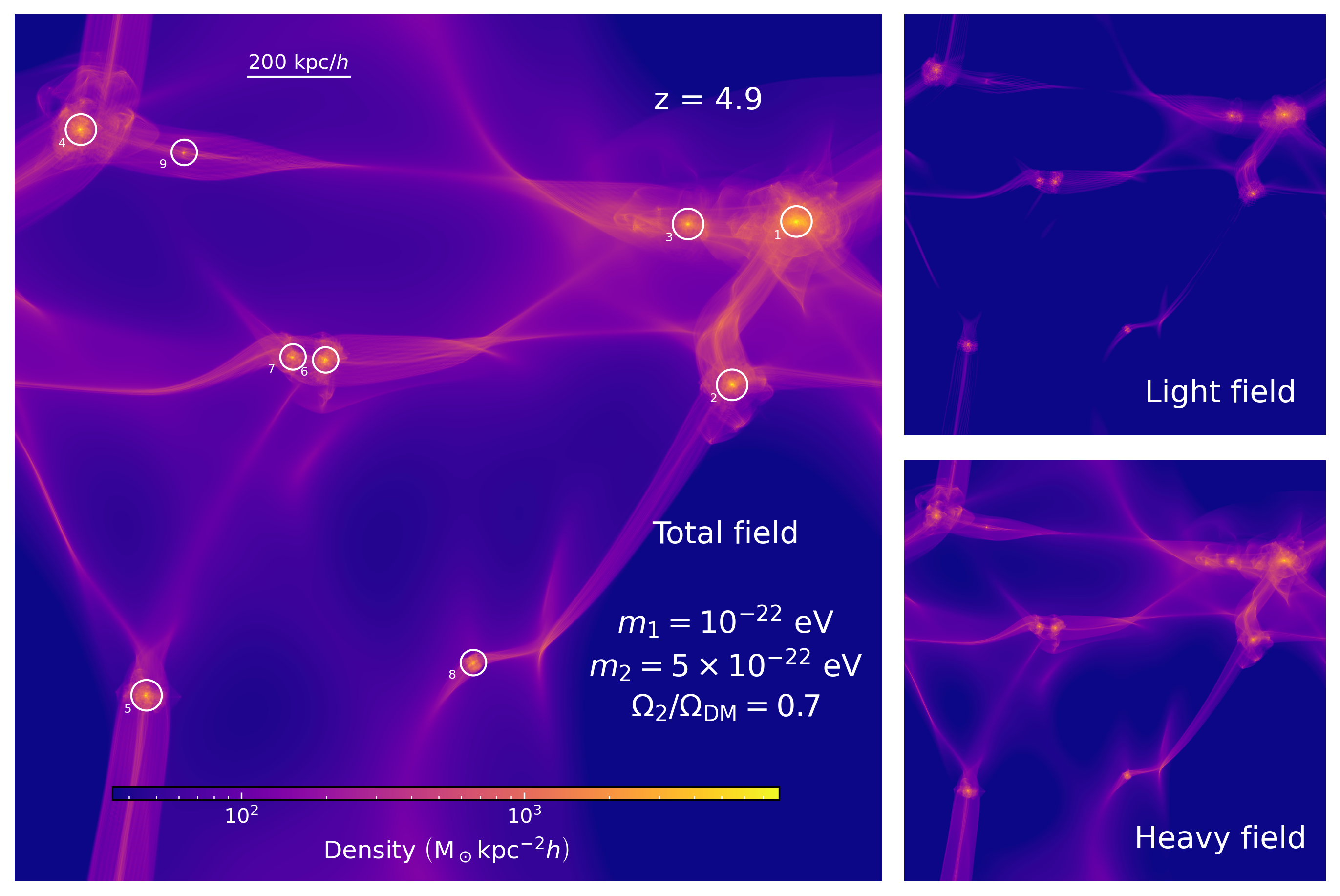}
\includegraphics[width=0.4\textwidth,height=4.75cm]{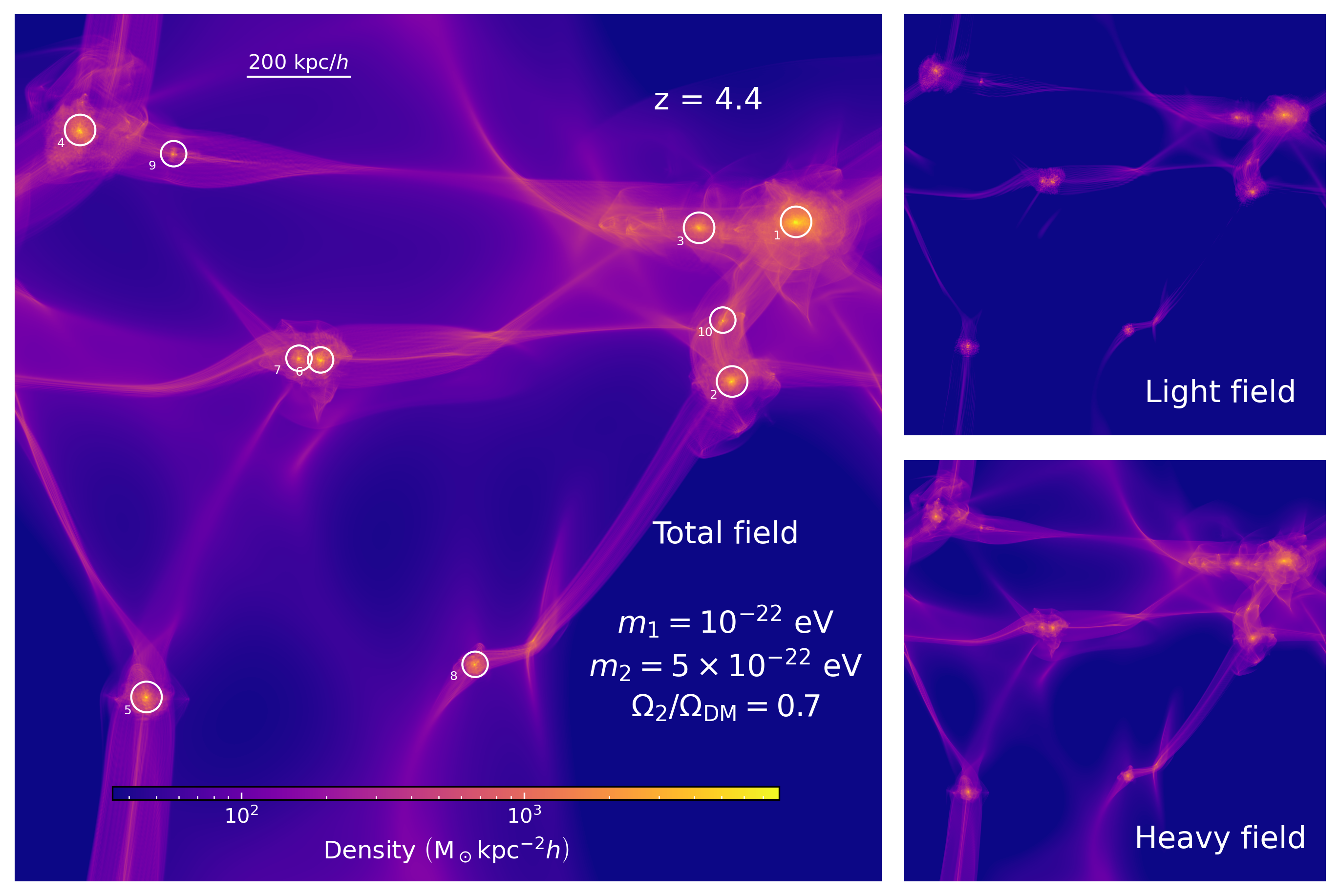}
\includegraphics[width=0.4\textwidth,height=4.75cm]{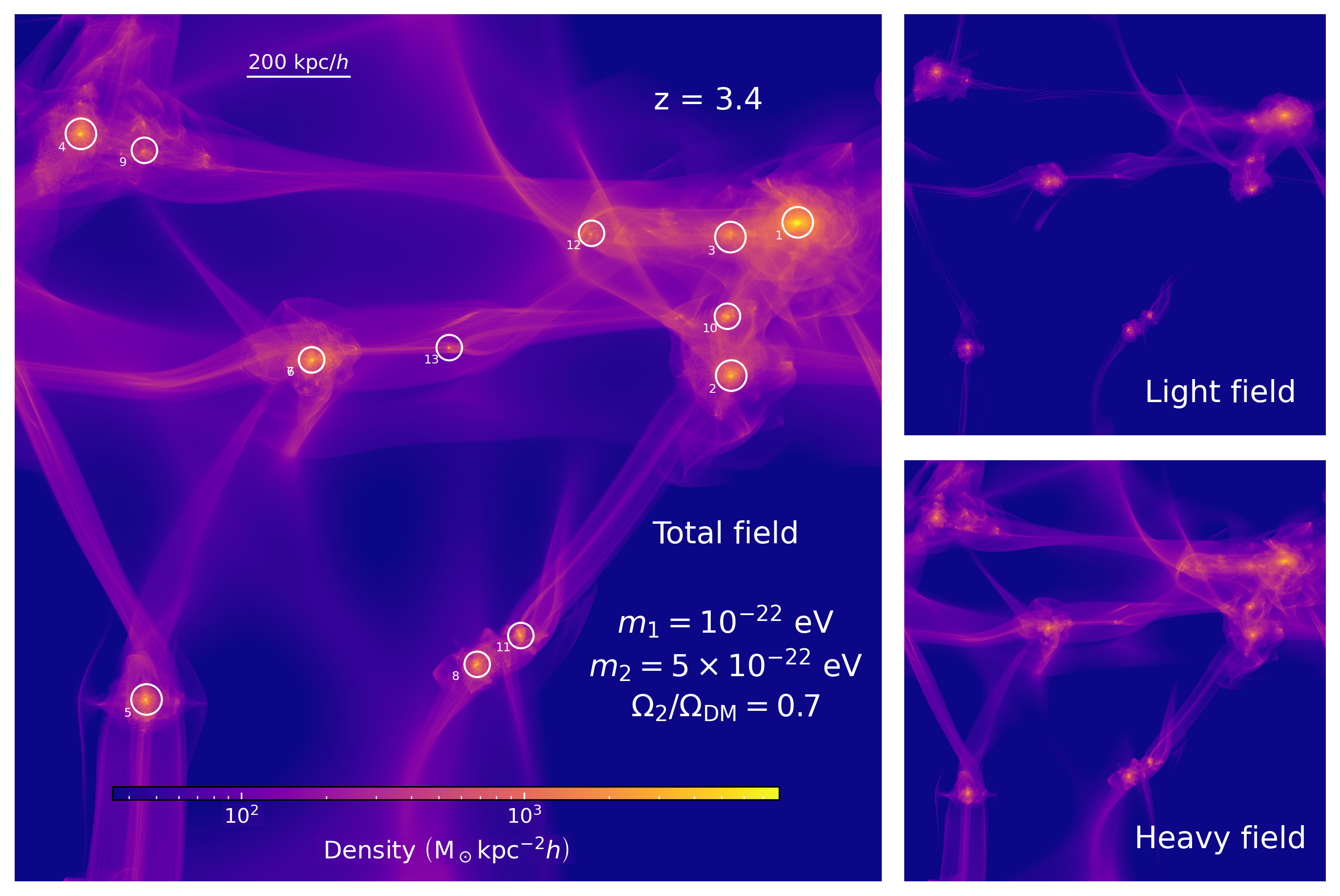}

   \caption{Projected densities of the total field within the simulation volume are shown at various redshifts. White circles highlight all haloes exhibiting a solitonic core structure at the corresponding redshift, as indicated in each panel. The labels next to each halo reflect their approximate formation sequence, with lower numbers representing earlier-forming haloes. Note that halos 6 and 7 have merged by the final timestep. }
   \label{Fig:Box}
\end{figure*}

\end{appendix}
\end{document}